\documentclass[twocolumn,aps,floatfix,superscriptaddress,preprintnumbers,amsmath,amssymb]{revtex4}

\usepackage{graphicx,dsfont}
\usepackage{color}

\begin{document}

\title{Formation of incommensurate long-range magnetic order in the  Dzyaloshinskii-Moriya antiferromagnet Ba$_2$CuGe$_2$O$_7$ studied by neutron diffraction}

\author{S. M\"uhlbauer}

\email{Sebastian.muehlbauer@frm2.tum.de}
\affiliation{Heinz Maier-Leibnitz Zentrum (MLZ), Technische Universit\"at M\"unchen, D-85748 Garching, Germany}

\author{G. Brandl}

\affiliation{J\"ulich Centre for Neutron Science (JCNS) at Heinz Maier-Leibnitz Zentrum (MLZ), Forschungszentrum J\"ulich GmbH, D-85748 Garching, Germany}
\affiliation{Physik Department E21, Technische Universit\"at M\"unchen, D-85748 Garching, Germany}

\author{M. M\aa nsson}

\affiliation{Materials Physics, KTH Royal Institute of Technology, Electrum 229, SE-16440 Stockholm Kista, Sweden}

\author{M. Garst}

\affiliation{Institut f\"ur Theoretische Physik, Technische Universit\"at Dresden, 01062 Dresden, Germany}

\date{\today}

\begin{abstract}
Neutron diffraction on a triple-axis spectrometer and a small-angle neutron scattering instrument is used to study the magnetic phase transition in tetragonal Ba$_2$CuGe$_2$O$_7$ at zero magnetic field. In addition to the incommensurate cycloidal antiferromagnetic (AFM) long-range order, we establish that weak incommensurate ferromagnetism (FM) also arises below the transition temperature $T_N$ identified by sharp Bragg peaks close to the $\Gamma$ point. The intensities of both the incommensurate AFM and FM Bragg peaks vanish abruptly at $T_N$ indicative of a weak first-order transition. Above $T_N$, evidence is presented that the magnetic intensity within the tetragonal $(a,b)$ plane is distributed on a ring in momentum space whose radius is determined by the incommensurate wavevector of the cycloidal order. We speculate that the associated soft fluctuations are at the origin of the weak first-order transition in the spirit of a scenario proposed by Brazovskii. 
\end{abstract}


\vskip2pc

\maketitle

\section{Introduction}
The spin-orbit coupling in non-centrosymmetric magnets leads to a Dzyaloshinskii-Moriya interaction (DMI) \cite{Dzyaloshinskii:58, Moriya:60} that favours incommensurate cycloidal or helical magnetic order. It is the competition between the symmetric exchange interaction and the DMI that generates a one-dimensional periodic modulation of the magnetization. In the limit of weak spin-orbit coupling, the associated wavelength becomes large or, equivalently, the size of its wavevector becomes small. At the same time, the magnetocrystalline anisotropies are, as a consequence, less effective in fixing the orientation of the magnetic order. 
Depending on the space group of the material, the energies of periodic magnetic textures might then be almost degenerate for orientations of their wavevectors that belong to manifolds in momentum space, e.g., a sphere.

Such a manifold of almost degenerate states can qualitatively influence the phase transition between the paramagnetic and the magnetically ordered phase. Upon approaching the phase transition from high temperatures, the paramagnons become soft on this manifold. The associated large phase space gives rise to a large entropy that might drive the phase transition first-order. This mechanism of a fluctuation-induced first-order transition was originally proposed by Brazovskii \cite{Brazovskii:1975wr} and theoretically discussed in various contexts, for example, for the theory of weak crystallization \cite{1987ZhETF..93.1110B}, liquid crystals \cite{1975JETP...42..497B,Swift:1976vp}, diblock copolymers \cite{Leibler:1980ti,Bates:1990hi}, pion condensation in nuclear matter \cite{1990PhR...192..179M}, and Bose-Einstein condensates \cite{Gopalakrishnan:2009gv}.

Recently, experimental evidence was presented by Janoschek  {\it et al.} \cite{Janosch:13} that such a Brazovskii scenario of a fluctuation-induced first-order transition is realized in the cubic helimagnet MnSi with paramagnons becoming soft on a sphere in momentum space. 
According to Brazovskii theory, the inverse correlation length $\kappa$ of the paramagnons 
will possess a characteristic temperature, $T$, dependence. It changes its curvature as the critical temperature is approached from a mean-field like behavior, $\partial_T^2 \kappa < 0$, at high $T$ to a regime, $\partial_T^2 \kappa > 0$, that is governed by strong fluctuations. Such a change in curvature was experimentally found in MnSi \cite{Janosch:13} in addition to the latent heat associated with the weak first-order transition \cite{Stishov:2007jf,2013PhRvL.110q7207B}. Subsequently, it was demonstrated that other cubic helimagnets with the same space group like Cu$_2$OSeO$_3$ \cite{PhysRevB.89.060401} and FeCo$_{1-x}$Si$_x$ \cite{2017arXiv170105448B} show similar behavior.

It is an interesting open question whether such a fluctuation-driven first-order transition might be also realized in magnetic systems with lower symmetry. A candidate in this regard is the non-centrosymmetric tetragonal antiferromagnet (AFM) Ba$_2$CuGe$_2$O$_7$ which crystallizes in the structure  $P{\overline 4}2_1m$ with lattice parameters $a=8.466$ \AA\, and $c=5.445$ \AA. Below $T_N\approx 3.05$\,K long-range order is realized with an incommensurate, cycloidal modulation of the AFM order parameter due to the DMI. The basic feature of Ba$_2$CuGe$_2$O$_7$ is a square arrangement of Cu$^{2+}$-ions in the $(a,b)$ plane. Nearest-neighbor AFM exchange along the diagonal of the $(a,b)$ plane is the dominant magnetic interaction ($J_{\|}\approx 0.96\,\rm {meV}$ per bond) \cite{Zheludev:99}. The coupling between adjacent planes is weak and ferromagnetic ($J_{\perp}\approx -0.026\,\rm {meV}$), essentially leading to a system of weakly coupled two-dimensional layers. A schematic depiction of the unit cell of Ba$_2$CuGe$_2$O$_7$ together with the orientation of the Dzyaloshinskii-Moriya vectors can be found in Ref.\,\cite{muehlbauer:12}. 

Numerous neutron scattering studies \cite{Zheludev:96, Zheludev:97, Zheludev:99, Zheludev:97b,Zheludev:97PB, Muehlbauer:11, muehlbauer:12} and theoretical considerations \cite{Zheludev:99,Bogdanov:02,Chovan:02,Chovan:13} have established a quite complete understanding of the AFM cycloidal order. It is characterized by a small incommensurate wavevector with length $k_{\rm cycl.} \equiv \sqrt{2} \xi$  r.l.u.~$\approx 0.028/$\AA~(1 r.l.u.$\, = 2\pi/a \approx 0.74/$\AA) and $\xi = 0.027$ which corresponds to a rotation of the AFM order by an angle 
of 9.7$^{\circ}$ per unit cell. The wavevector is located within the $(a,b)$ plane, and weak magnetocrystalline anisotropies favour an alignment along a $[1,\pm1,0]$ direction with the magnetic moments confined to the $(1,\mp1,0)$ planes. This results in two domains that are equally populated in case of zero-field cooling, giving rise to magnetic satellite reflections in reciprocal space at $(1\pm\xi,\pm\xi,0)$ and $(1\pm\xi,\mp\xi,0)$ centered around (1,0,0), which corresponds to the N\'{e}el point for Ba$_2$CuGe$_2$O$_7$.

The transition between the zero-field paramagnet and the cycloidal AFM order in Ba$_2$CuGe$_2$O$_7$ is weakly first-order. Similar to MnSi \cite{Stishov:2007jf,Bauer:10}, the specific heat shows a sharp latent heat which sits on top of a broader peak \cite{muehlbauer:12}. The origin of the first-order transition, however, has remained unclear so far. A mean-field approximation of the proposed effective theory \cite{Bogdanov:02,Chovan:02, Chovan:13} predicts a continuous second-order transition as a function of temperature suggesting that the first-order character is induced by fluctuations. 

In the present work, we performed neutron diffraction experiments on Ba$_2$CuGe$_2$O$_7$ using both a triple-axis spectrometer (TAS) and a small-angle neutron scattering (SANS) instrument. Whereas the former is ideally suited to investigate incommensurate AFM order around the finite N\'eel wavevector in reciprocal space, the latter experiment allows us to study FM compondents of the magnetic order close to the $\Gamma$ point, i.e., zero wavevector. It was pointed out in the theoretical works of Refs.~\cite{Bogdanov:02,Chovan:02} that there exists a Dzyaloshinskii vector pointing along the $c$ axis whose sign alternates from bond to bond giving rise to weak ferromagnetism (FM), see also the discussion in Ref.~\cite{muehlbauer:12}. As a consequence, the formation of the AFM cycloid is expected to be accompanied by the emergence of weak incommensurate FM order with the same wavevector. Our SANS experiment confirms this theoretical prediction. 

The main focus of this work are, however, the critical magnetic fluctuations in Ba$_2$CuGe$_2$O$_7$. We demonstrate that close to the critical temperature $T_N$ magnetic intensity indeed accumulates on a manifold  in momentum space. With the employed scattering geometry, we were limited to wavevectors within the crystallographic $(a,b)$ plane where this manifold corresponds to a ring with radius $k_{\rm cycl.}$.  This suggests that magnetocrystalline anisotropies, that eventually pin the cycloidal propagation direction below $T_N$, are comparatively small. These observations fulfil in principle the preconditions for a fluctuation-induced first-order transition of Brazovskii type. However, the temperature dependence of the inverse correlation length, $\kappa(T)$, of the critical paramagnons is not in agreement with the predictions of Brazovskii theory. Instead, we find that $\kappa(T)$ does not change its curvature but it is always convex, $\partial_T^2 \kappa > 0$, in an extended temperature range above $T_N$. We attribute the convexity of $\kappa(T)$ to the effective low dimensionality of fluctuations due to the weak coupling $J_\perp$ between the layers.

The structure of the paper is as follows. In section \ref{sec:theory} we first review the theory of cycloidal antiferromagnetic order in Ba$_2$CuGe$_2$O$_7$ and discuss its predictions for the differential cross section for neutron diffraction. In section \ref{sec:ExpMethods} we describe the details of the employed experimental methods. The experimental results and their analysis are presented in section \ref{sec:neutrondiff}. Finally, in section \ref{sec:summary} we close with a summary and a discussion.

\section{Theory}
\label{sec:theory}

\subsection{Effective theory for cycloidal AFM order}
We first review the theoretical description of the AFM order in Ba$_2$CuGe$_2$O$_7$ following Chovan {\it et al.} \cite{Chovan:02}. At zero magnetic field, the orientation of the staggered magnetization specified by the unit vector $\hat n$ is governed by the theory $\mathcal{L} = \mathcal{L}_0+ \mathcal{L}_{\rm m.a.}$ where the first term reads
\begin{align} \label{theory}
\mathcal{L}_0  = \frac{1}{2} \Big[&
(\partial_x \hat n)^2 + (\partial_y \hat n)^2  +  K_z \hat n_z^2 \\\nonumber
&
- 2 Q (\hat e_y \times \hat n) \partial_x \hat n 
-2 Q (\hat e_x \times \hat n) \partial_y \hat n\Big].
\end{align}
Here, $Q> 0$ derives from the DMI and favours a modulated AFM texture with a finite wavevector. We treat the uniaxial anisotropy $K_z$ as a free parameter. However, it was pointed out by Kaplan \cite{Kaplan1983} and Shekhtman, Aharony, and Entin-Wohlman \cite{Shekhtman1993} that $K_z$ might be related to the DMI, and in this so-called KSEA limit the uniaxial anisotropy  can be identified with $K_z = Q^2$. 
The derivatives only act within the $(x,y)$ plane, i.e., the crystallographic $(a,b)$ plane. A dispersion along the crystallographic $c$ axis is here neglected. On the level of Eq.~\eqref{theory}, the theory possesses a U(1) symmetry representing the combined rotation of real and order parameter space around the $z$ axis. 

This symmetry gets explicitly broken by in-plane magnetocrystalline anisotropies contained in $\mathcal{L}_{\rm m.a.}$. There are various contributions to $\mathcal{L}_{\rm m.a.}$ but we just confine ourselves to one representative term 
\begin{align} \label{cub}
\mathcal{L}_{\rm m.a.}  = \frac{\lambda}{2}   \hat n (\partial_x^4 + \partial_y^4 - \frac{1}{2}(\partial_x^2 + \partial_y^2)^2) \hat n
\end{align}
that fixes the orientation of the wavevector of the magnetic texture within the ordered phase. 
The last term in Eq.~\eqref{cub}, that is isotropic, has been introduced for later convenience.
 
Below the critical temperature $T_N$, a flat spiral is realized along a certain direction $\hat e_1$ within the $(x,y)$ plane, i.e., the crystallographic $(a,b)$ plane 
\begin{align} \label{AFMcycloid}
\hat n(r_1) = \hat e_1\,\sin \phi(r_1)  +  \hat e_3\,\cos \phi(r_1)
\end{align}
where $r_1 = \vec r \hat e_1$ with the orthonormal vectors $\hat e_i \hat e_j = \delta_{ij}$, $i,j=1,2,3$. The third component coincides with the crystallographic $c$ axis,  $\hat e_3 = \hat z$. The field $\phi$ satisfies at $\lambda = 0$ a sine-Gordon equation so that the spiral is not harmonic even at zero magnetic field. Due to the U(1) rotation symmetry of $\mathcal{L}_0$ the direction of $\hat e_1$ remains undetermined on the level of Eq.~\eqref{theory}. This degeneracy is explicitly broken by the weak in-plane anisotropies of Eq.~\eqref{cub}.
A finite  in-plane anisotropy $\lambda > 0$ favours an orientation $\hat e_1$ along crystallographic $[\pm1,\pm1,0]$ or $[\pm1,\mp1,0]$ directions consistent with experimental observations. 

\begin{widetext}

\subsection{Critical fluctuations of AFM order}

Above the critical temperature $T_N$, fluctuations of the AFM order parameter 
are governed by the staggered susceptibility matrix that derives from Eqs.~\eqref{theory} and \eqref{cub}
\begin{align} \label{Sus}
\chi^{-1}_{\rm AFM}({\bf q}) = \left(\begin{array}{ccc}
r + q^2 + \lambda (q_x^4 + q_y^4 - \frac{1}{2} q^4) & 0 &  i 2 Q q_x 
\\
0 & r + q^2 + \lambda (q_x^4 + q_y^4 - \frac{1}{2} q^4) & -i 2 Q q_y
\\
-i 2 Q q_x &  i 2 Q q_y & r + K_z + q^2 + \lambda (q_x^4 + q_y^4 - \frac{1}{2} q^4)
\end{array}
\right)
\end{align}
where $q^2 = {\bf q}^{2}$. The wavevector ${\bf q} = (q_x,q_y)$ is here measured with respect to one of the 
N\'eel vectors, i.e., $(2\pi/a,0,0)$ or $(0, 2\pi/a,0)$. In the following, we concentrate on the domain with 
the N\'eel vector ${\bf G} = (2\pi/a,0,0)$. The parameter $r$ tunes the distance to the phase transition.
The lowest eigenvalue of Eq.~\eqref{Sus} reads
\begin{align}
\varepsilon({\bf q}) = r + \frac{K_z}{2} - \frac{1}{2}\sqrt{K_z^2 + 16 q^2 Q^2} +q^2 + \lambda (q_x^4 + q_y^4- \frac{1}{2} q^4).
\end{align}
Without  in-plane magnetocrystalline anisotropies $\lambda = 0$, it attains its minimum on a ring within the $(x,y)$ plane of momentum space. As we neglected the dispersion along the $z$ axis, this ring extends along the third axis forming a cylinder. A finite in-plane anisotropy with $\lambda>0$ results in modulations of intensity on the ring with minima along directions $[110]$, $[\bar{1}10]$, $[1\bar{1}0]$, and $[\bar{1}\bar{1}0]$. The radius of the ring is given by 
\begin{align}
k_{\rm cycl.} = \sqrt{Q^2 - \frac{K^2_z}{16 Q^2}}.
\end{align}
The minimum value of the eigenvalue $\varepsilon({\bf q})$ reads
\begin{align} \label{kappa}
\kappa^2 \equiv r - \Big(Q - \frac{K_z}{4Q}\Big)^2, 
\end{align}
which can be identified with the square of the inverse correlation length $\kappa$ of cycloidal AFM order. 

For neutron diffraction with respect to the N\'eel vector probing AFM order the differential cross section is proportional to the trace tr$\{(\mathds{1} - \hat G \hat G^T) \chi_{\rm AFM}\}$ with $\hat G = (1,0,0)$. It reads explicitly
%
\begin{equation}
\frac{d\sigma({\bf q})}{d\Omega}\Big|_{\rm AFM} = 
A k_B T \frac{(r+q^2)(2(r+q^2) + K_z) - 4 q^2 Q^2 \cos^2 \phi}{\Big(r+q^2\Big)
\Big(r+q^2 + \frac{K_z}{2} + \frac{1}{2}\sqrt{K_z^2 + 16 q^2 Q^2}\Big)
\Big(r+q^2 + \frac{K_z}{2} - \frac{1}{2}\sqrt{K_z^2 + 16 q^2 Q^2} + \frac{\lambda q^4}{4}(1+\cos(4\phi)) \Big)}
%
\label{eq1}
\end{equation}
where $A$ is a constant proportionality factor, $k_B$ is the Boltzmann constant and $T$ is the temperature.
The reduced momentum transfer ${\bf q}=q(\cos\phi,\sin\phi)$ is located within the $(a,b)$ plane. In the above expression, we anticipated that the  in-plane anisotropies are effectively small, and we kept $\lambda$ only for the lowest eigenvalue appearing in the denominator. Whereas the dependence on the angle $\phi$ in the denominator is induced by the anisotropy $\lambda$, the dependence in the numerator 
derives from the dipolar interactions that effectively project the magnetic moments onto a subspace perpendicular to the N\'eel vector $\hat G$.

\subsection{Critical fluctuations of FM order}

According to Refs.~\cite{Bogdanov:02,Chovan:02}, the AFM order in Ba$_2$CuGe$_2$O$_7$ will induce weak ferromagnetism due to the presence of a bond-alternating Dzyaloshinskii vector pointing along the $z$ axis. The induced ferromagnetic component  reads
\begin{align} \label{inducedFM}
\delta \vec m = d\, \hat n \times \hat z
\end{align}
with a small proportionality constant $d$. The AFM cycloid of Eq.~\eqref{AFMcycloid} therefore gives rise to an incommensurate FM density-wave with the same wavevector where the FM order parameter $\delta \vec m$ vanishes whenever $\hat n$ is aligned along $\hat z$. Chovan {\it et al.} \cite{Chovan:02} found that there exists an additional contribution $\delta \vec m \propto \partial_x \hat n$ with the FM polarization being proportional to the derivative of $\hat n$ along the $x$ direction, i.e., the direction of the N\'eel vector. As our experimental data does not allow to discriminate between the two contributions, we will limit ourselves here to Eq.~\eqref{inducedFM}. According to the relation Eq.~\eqref{inducedFM}, the critical fluctuation of the FM and AFM order parameter are related by $(\chi_{\rm FM})_{ij} \propto \varepsilon_{inz} \varepsilon_{jmz} (\chi_{\rm AFM})_{nm}$.

The differential cross section for elastic neutron scattering with momentum transfer ${\bf q}=q(\cos\phi,\sin\phi)$ but with respect to zero wavevector is then proportional to tr$\{(\mathds{1} - \hat q \hat q^T)\chi_{\rm FM}({\bf q})\}$ where $\hat q = {\bf q}/q$. It reads
\begin{equation}
\frac{d\sigma({\bf q})}{d\Omega}\Big|_{\rm FM} = B k_B T 
\frac{(r+q^2)(r+q^2 + K_z) - 4 q^2 Q^2 \sin^2(2 \phi)}
{\Big(r+q^2\Big)
\Big(r+q^2 + \frac{K_z}{2} + \frac{1}{2}\sqrt{K_z^2 + 16 q^2 Q^2}\Big)
\Big(r+q^2 + \frac{K_z}{2} - \frac{1}{2}\sqrt{K_z^2 + 16 q^2 Q^2} + \frac{\lambda q^4}{4}(1+\cos(4\phi)) \Big)}
\label{DCS-FM}
\end{equation}
where $B$ is another constant proportionality factor, and we employed the same approximation concerning $\lambda$ as in Eq.~\eqref{eq1}.

\end{widetext}

We will use Eqs.~\eqref{eq1} and \eqref{DCS-FM} for fitting our neutron diffraction data of the AFM and FM fluctuations assuming that the temperature dependence of $A, B, Q, K_z$ and $\lambda$ is negligible close to $T_N$. The strong temperature dependence of the scattering data is only accounted for by the tuning parameter $r=r(T)$ or, alternatively, by the inverse correlation length $\kappa(T)$, see Eq.~\eqref{kappa}. 

\section{Experimental Methods}
\label{sec:ExpMethods}

For our neutron diffraction measurements a single crystal of Ba$_2$CuGe$_2$O$_7$ with a diameter of $\sim 5\,{\rm mm}$ and a length of $\sim 19\,{\rm mm}$ has been used. This crystal has already been utilized in several neutron diffraction experiments before \cite{Muehlbauer:11, muehlbauer:12, Muehlbauer:15}. A detailed description of the growth recipe and annealing process of the sample is given in Ref.\,\cite{muehlbauer:12}. The experiments have been performed on the triple axis spectrometer TASP at PSI, Villigen \cite{semadeni:01, janoschek:07} and the small-angle neutron scattering instrument SANS-1 at MLZ, Garching \cite{SANS-1, Muehlbauer:16}.

\begin{figure*}
\begin{center}
\includegraphics[width=0.99\textwidth]{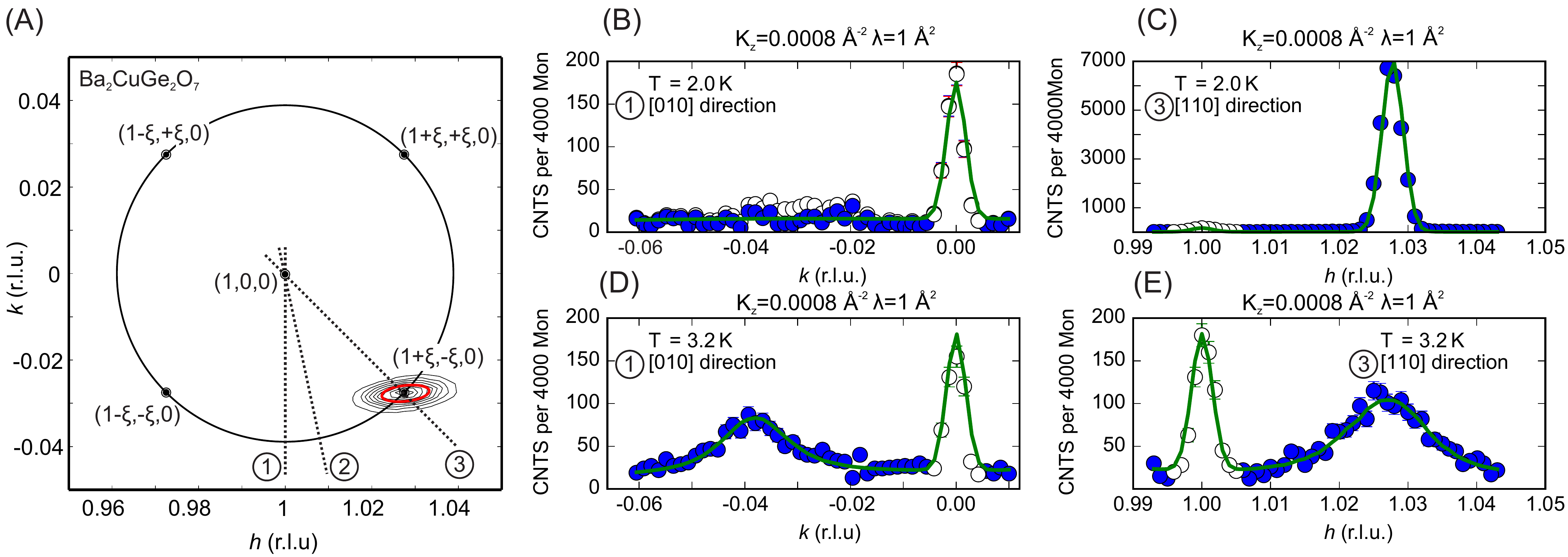}
\caption{
Incommensurate AFM order and critical fluctuations in Ba$_2$CuGe$_2$O$_7$ as measured with the triple-axis spectrometer (TAS). Panel (A): Map of the reciprocal space around the (1,0,0) position. Four incommensurate satellite reflections associated with the cycloidal AFM order arise at 
($1\pm\xi$,$\pm\xi$,0) and ($1\pm\xi$,$\mp\xi$,0). The contours around ($1+\xi$,$-\xi$,0) indicate the experimental resolution ellipsoid of the TAS instrument in the $(a,b)$-plane. 
The dotted lines denoted with \textcircled{\scriptsize 1}, \textcircled{\scriptsize 2} and \textcircled{\scriptsize 3} indicate the orientation of the different scans with respect to the incommensurate satellites and the ring of intensity. Panels (B-E): Typical data obtained on scans \textcircled{\scriptsize 1} and \textcircled{\scriptsize 3} for a temperature of 2\,K (below the transition temperature $T_N$, (B) and (C)) and 3.2\,K (above $T_N$, (D) and (E)). Open data points indicate spurious intensity observed $(i)$ due to higher order intensity at the commensurate (1,0,0) position and $(ii)$ due to broad wings of the resolution ellipsoid. The green line represents the global fit of the data to Eq.~\eqref{eq1} with on particular set of parameters $K_z = 0.0008/$\AA$^2$ and $\lambda = 1 $\AA$^2$. }
\label{Fig_1}
\vspace{-0.00\textwidth}
\end{center}
\end{figure*}

On TASP, the sample has been mounted inside a standard orange type cryostat with its [0,0,1] crystalline direction vertically aligned. A base temperature of 1.55\,{\rm K} with a stability of $\pm$0.01\,{\rm K} is reached. The alignment of the sample was performed using the (1,1,0) and (2,0,0) nuclear reflections. TASP was configured in "2-axis mode" with no energy analysis of the scattered neutrons. The scattered intensity hence represents an integration over all neutron energies \footnote{In fact, the analyzer was not removed but rotated such that the neutrons transmitted the analyzer crystals under an angle of 90$^{\circ}$ leading to a reduced transmission of 85\%}. A tight collimation of 20' was used in front of the monochromator and sample, but no collimator was placed between sample and detector. The incident neutron energy was adjusted to $E_i=3.5\,{\rm meV}$.

For the experiment on SANS-1 the sample was aligned with the $c$ axis parallel to the incident neutron beam and a [1,0,0] axis vertical. The sample was cooled with a three stage dry closed cycle cryostat to a base temperature of 1.5\,K with a temperature stability of $\pm$0.003\,{\rm K}. Neutrons with a wavelength of ${\rm 5\, \AA}$ were collimated over a distance of 16\,m. A sample-detector distance of 12\,m has been used. The background signal was determined for a temperature well above $T_N$ and subtracted from the data.

\section{Neutron diffraction close to the magnetic transition in  ${\rm \bf Ba_2CuGe_2O_7}$}
\label{sec:neutrondiff}

In the following, we present neutron diffraction data close to the magnetic phase transition in Ba$_2$CuGe$_2$O$_7$ at zero field. We first discuss the results obtained with the help of a triple-axis spectrometer (TAS) where we focussed on the AFM fluctuations close to the N\'eel wavevector. Afterwards we present our small-angle neutron scattering (SANS) data that is sensitive to the FM fluctuations close to zero wavevector. The latter experiment in particular confirms that the formation of cycloidal AFM order also induces weak incommensurate FM order with the same wavevector.

\subsection{AFM order probed with TAS}

\begin{figure}
\includegraphics[width=0.45\textwidth]{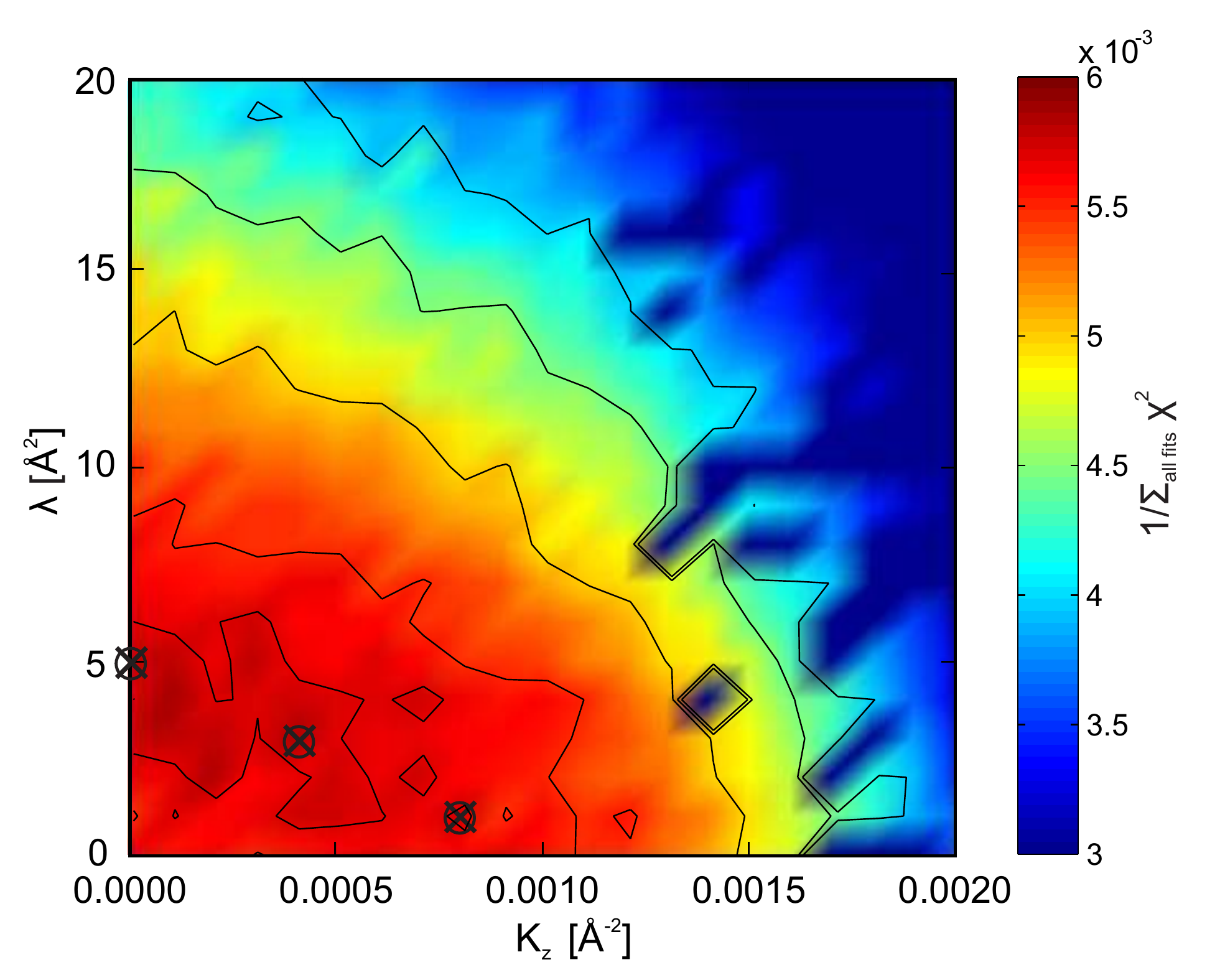}
\caption{Fit quality of the TAS data: map of $1/\sum_{\rm all\, fits}{\chi^2}$ as a function of $K_z$ and $\lambda$. The marks indicate combinations of $(K_z,\lambda)$ plotted in Fig.\,\,\ref{Fig_4}.}
\label{Fig_2}
\vspace{-0.00\textwidth}
\end{figure}

\begin{figure*}
\begin{center}
\includegraphics[width=0.99\textwidth]{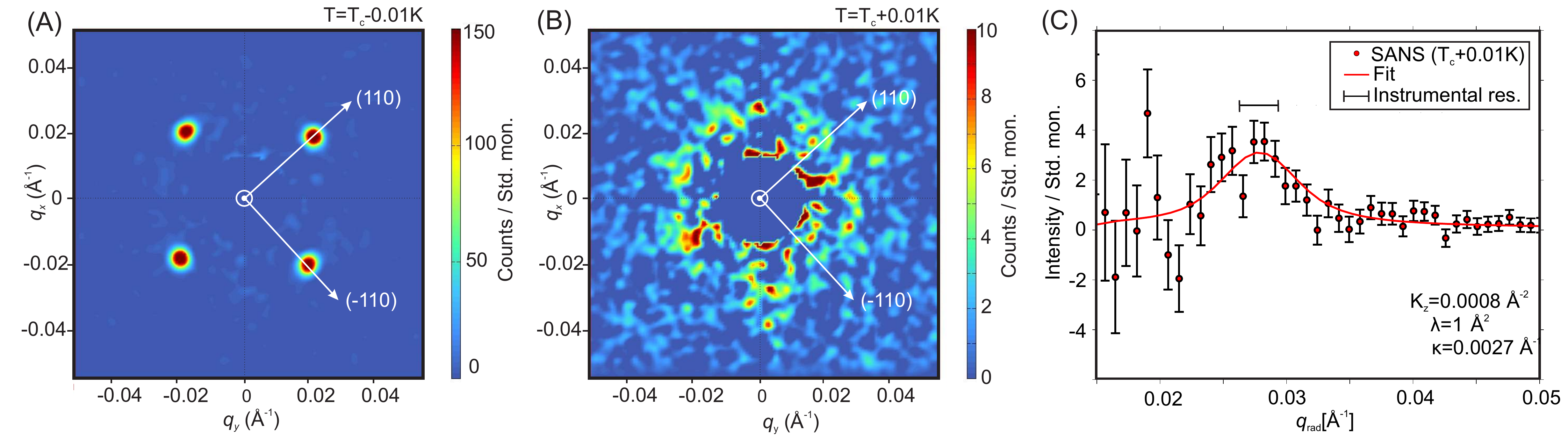}
\caption{Incommensurate FM order and critical fluctuations in Ba$_2$CuGe$_2$O$_7$ as measured with small-angle neutron scattering (SANS). Panel (A): Typical SANS diffraction pattern for $T<T_N$ revealing incommensurate FM order. The diffraction pattern is obtained by a sum over a rocking scan with respect to the vertical axis. Pixels around the direct beam are masked electronically. The background has been recorded for temperatures well above $T_N$ and is subtracted. Panel (B) shows corresponding data at temperatures closely above $T_N$ where  a weak ring feature is observed. Panel (C) shows radially integrated data of the scattering pattern of panel (B) including the instrumental resolution and the fit to theory, see text.}
\label{Fig_3}
\vspace{-0.00\textwidth}
\end{center}
\end{figure*}

On the TAS instrument, the incommensurate order arising around the forbidden nuclear reflection at (1,0,0) has been investigated with the help of three scans denoted by \textcircled{\footnotesize 1}, \textcircled{\footnotesize 2} and \textcircled{\footnotesize 3}. Panel (A) of Fig.~\ref{Fig_1} shows a map of the reciprocal space and illustrates the direction of the three different scans. The instrumental resolution has been determined with a dense mesh of points covering the incommensurate satellite at ($1+\xi$,$-\xi$,0) at the lowest temperature available, as indicated by the contour lines in panel (A), which was fitted to a 2D-Gaussian function.
Each scan was performed as function of increasing temperature between 1.55\,\,K and 4\,\,K with a minimum spacing of 0.02\,K. Scans denoted with \textcircled{\footnotesize 1} are oriented along the [0,1,0] direction. Scans denoted with \textcircled{\footnotesize 2} cross the ring of intensity observed at temperatures above $T_N$ such that the short axis of the resolution ellipsoid is oriented perpendicular to the ring.  Scans denoted with \textcircled{\footnotesize 3} point along the diagonal direction towards ($1+\xi$,$-\xi$,0). As the scans only cover a small portion of reciprocal space, the resolution of the TAS was assumed to be constant within this region. 

Typical experimental data of scans along \textcircled{\footnotesize 1} and \textcircled{\footnotesize 3} are shown in Fig.~\ref{Fig_1} panels (B) to (E). The open data points indicate spurious intensity: (i) The intensity observed at the forbidden, commensurate position at (1,0,0) is due to multiple nuclear scattering and has been found to be temperature independent. (ii) A shallow spurious peak at $(1,-0.03,0)$ (panel (D)), which is attributed to wings of the resolution ellipsoid. We attribute the blue-filled data points to magnetic scattering. Scans in the upper row  panels (B) and (C) are recorded below the transition temperature $T_N \approx 3.05$ K. In the scan \textcircled{\footnotesize 3} at $T=2$ K, a resolution limited Gaussian peak at ($1+\xi$,$-\xi$,0) associated with the cycloidal AFM structure is clearly observed. In contrast, there is no magnetic intensity observed at the same temperature in the scan \textcircled{\footnotesize 1} apart from the shallow spurious peak at $(1,-0.03,0)$. Scans taken for temperatures slightly above $T_N$ are shown in the lower row of Fig.~\ref{Fig_1}, panels (D) and (E). Here, a clearly broadened peak with Lorentzian-like lineshape emerges at ($1,-\sqrt{2}\xi,0)$  in the scan \textcircled{\footnotesize 1} at $T=3.2$ K. In the scan \textcircled{\footnotesize 3} the sharp incommensurate Bragg peak is now replaced by a largely broadened peak. Both broad peaks are consistent with a ring of intensity centered at (1,0,0) with radius $k_{\rm cycl.} = \sqrt{2}\xi$ r.l.u..

The TAS data has been analysed using a model that includes the following ingredients: For temperatures below $T_N$: sharp resolution limited peaks at the positions of the incommensurate satellite reflections ($1\pm\xi$,$\pm\xi$,0) and ($1\pm\xi$,$\mp\xi$,0). For all temperatures: $(i)$ spurious intensity at the forbidden nuclear position at (1,0,0) and at $(1,-0.03,0)$, and $(ii)$ a ring of intensity around (1,0,0), described by Eq.~\eqref{eq1}, where the angle $\phi=0^{\circ}$ corresponds to the [1,0,0] direction. Additionally, $(iv)$ a temperature independent constant background has been subtracted.

As a first step in the analysis, the shallow spurious peaks found in scans \textcircled{\footnotesize 1} and \textcircled{\footnotesize 2} below the transition temperature $T_N$ are fitted for the lowest temperature measured ($T=1.5$ K). The spurious intensity in each case is found at $(h,k,0)$ with $k \approx -0.03$ and $|k| < \sqrt{2}\xi$. Most likely, it is caused by the shallow wings of the incommensurate peak at $(1+\xi,-\xi,0)$ that are not captured by the 2D-gaussian fit of the resolution ellipsoid. The spurious peaks are subtracted from the data assuming the same temperature dependence as found for the main satellite peaks. In the next step, the data of both scans \textcircled{\footnotesize 1} and \textcircled{\footnotesize 2} are fitted to Eq.~\eqref{eq1} with the angle $\phi$ fixed to the corresponding values of both scans. After accounting for the instrumental resultion, the fits were first performed with a free scaling factor $A$ for each temperature independently. Note that Eq.~\eqref{eq1} is only valid for $T>T_N$ but is used here phenomenologically also for $T<T_N$ to describe the remaining magnetic intensity on the ring away from the Bragg peaks. The results of these fits are used to fit the intensity of the incommensurate satellite peak for each temperature from the scan \textcircled{\footnotesize 3}. The fit is finally refined with a common value of $A$ for all temperatures. For the fits, we have fixed the value of $k_{\rm cycl.} = 0.028/$\AA, which leaves us with three fitting parameters, $K_z$, $\lambda$ and $\kappa$, apart from the scaling factor $A$. The KSEA limit $K_z = Q^2$ then corresponds to the value $K_z|_{\rm KSEA} \approx 0.0008/$\AA$^2$. The fitting procedure is systematically repeated for different combinations of the parameters $K_z$ and $\lambda$ ranging from $0$ to $0.002/$\AA$^{2}$ for $K_z$ and from $0$ to $20$ \AA$^2$ for $\lambda$. 

To quantify the quality of the fits, $1/\sum_{\rm all\, fits}{\chi^2}$ is plotted as a figure of merit (FOM) as a function of the parameters $K_z$ and $\lambda$ in Fig.~\ref{Fig_2}. A shallow ridge is observed for combinations of $\lambda$ and $K_z$ along a line from $(K_z=0,\lambda=5$\AA$^{2})$ to $(K_z=0.001/$\AA$^{2},\lambda=0)$. Close to the ridge observed for the FOM, an excellent agreement is observed. A typical global fit to all TAS data is included in all panels of Fig.~\ref{Fig_1}, panels (B) to (E). Here, one particular set of parameters has been chosen $(K_z=0.0008/$\AA$^{2},\lambda=1$\AA$^{2})$ (solid green line), that is consistent with the KSEA limit. It shows an excellent overall agreement with the measurement data.

\subsection{FM order probed with SANS}
\label{sans}

We now turn to a discussion of the SANS data for Ba$_2$CuGe$_2$O$_7$ that probe the FM fluctuations close to zero wavevector. Panel (A) of Fig.~\ref{Fig_3} shows typical scattering data for $T<T_N$. Four diffraction peaks indicative of two domains of incommensurate FM order are observed. The complete diffraction pattern is obtained by a sum over a rocking scan with respect to the vertical axis. Below $T_N$, the width of the peaks are limited by the instrumental resolution. The background is measured well above $T_N$ and subtracted accordingly. The periodicity of the incommensurate wavevector coincides with the one observed for the AFM cycloid. 

Panel (B) of Fig.~\ref{Fig_3} illustrates that above $T_N$ the sharp diffraction spots give way to a weak ring feature. At the same time, the intensity drops dramatically. For the analysis, the diffraction patterns have been radially integrated; panel (C) shows typical data corresponding to the diffraction pattern of panel (B). A weak, ring-like feature is observed. The ring is fitted to Eq.~\eqref{DCS-FM} after accounting for the instrumental resolution, which is determined with the help of a fit to the resolution-limited peak below $T_N$. We have chosen the parameters $K_z=0.0008/$\AA$^2$ and $\lambda=1$ \AA$^{2}$, that were obtained from the description of the TAS data.

\subsection{Results of neutron diffraction experiments}

From the fit of the TAS data to Eq.~\eqref{eq1} we obtained the temperature dependence of the intensity of the incommensurate peaks associated with the cycloidal AFM structure, the intensity of the ring found around the N\'eel point (1,0,0) with radius $\sqrt{2}\xi$ and the inverse correlation length $\kappa$. 

\begin{figure}
\includegraphics[width=0.42\textwidth]{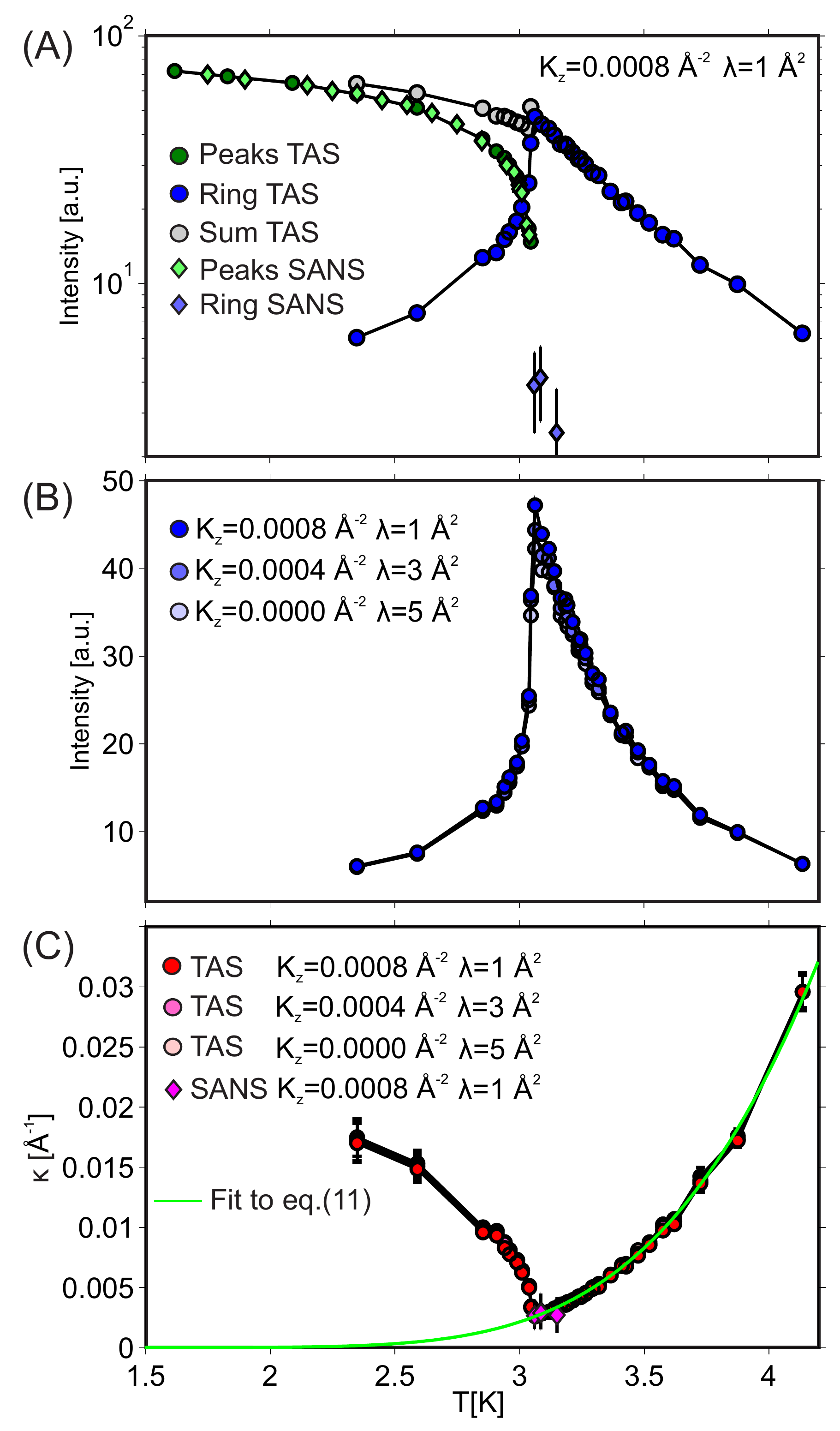}
\caption{Temperature dependence of the intensity and the inverse correlation length $\kappa$ from the TAS and SANS experiments. Panel (A) depicts the integrated intensity of the incommensurate Bragg peaks, including both domains, as a function of temperature below $T_N$ (green symbols) for the TAS (dots) and the SANS (diamonds) experiment. Blue symbols indicate the integrated intensity over the complete the ring for  $(K_z=0.0008/$\AA$, \lambda=1$ \AA$^{2})$ for both TAS (dots) and SANS (diamonds) data. Grey symbols indicate the sum of both ring and peaks. Panel (B) shows the intensity of the ring as function of temperature as obtained from the TAS experiment for different combinations of $K_z$ and $\lambda$. Panel (C) shows the temperature evolution of $\kappa$ for different combinations of $K_z$ and $\lambda$ for both SANS and TAS data.The green line is a fit to the exponential function of Eq.~\eqref{FitKappa}. The data points in panel (B) and (C) for the given values of $K_z$ and $\lambda$ are essentially equivalent illustrating their equal FOM, see Fig.~\ref{Fig_2}.}
\label{Fig_4}
\end{figure}

Our fit hence allows for a quantitative comparison of the intensity of the sum of both cycloidal AF domains versus the total intensity contained in the ring. The results are shown in panels (A), (B) and (C) of Fig.~\ref{Fig_4} for three different combinations of $K_z$ and $\lambda$ as indicated by the marks in the map of the FOM in Fig.~\ref{Fig_2} \footnote{ Note, that for panel (B) and (C) datapoints with different combinations of $K_z$ and $\lambda$ are essentially overlapping almost.}. The fit of the SANS data to Eq.~\eqref{DCS-FM}, which contains information about the FM fluctuations, yields in addition the intensity of the Bragg peaks of the incommensurate FM order. Due to the weak intensity of the ring in the SANS data only few points could be fitted reliably resulting in a single value of $\kappa$ for the critical FM fluctuations close to $T_N$.

Panel (A) of Fig.~\ref{Fig_4} depicts the temperature dependence of the intensity associated with the incommensurate satellite reflection (sum of both domains, green symbols) obtained from a fit with $(K_z=0.0008/$\AA$^2,\lambda=1$\AA$^2)$ together with the integrated intensity of the ring (blue symbols) and, for the TAS data, their sum (grey points). Note, that Fig.\,\,\ref{Fig_2} only representy cuts.
The integrated intensity of the Bragg peaks from the SANS data was scaled to the corresponding intensity of the TAS experiment yielding consistent results for the $T$ dependence. 

The same scaling factor suggests that the intensity associated with the ring as observed in SANS (blue diamonds in panel (A)) is, however, substantially lower than the one of the TAS data. We attribute this apparent discrepancy to the different integration range of the two experiments along the tetragonal axis perpendicular to the ring. While the TAS instrument accepts a vertical divergence of $4-5^{\circ}$ due to the focusing monochromator, the SANS instrument only accepts an extremely good divergence of $0.3^{\circ}$. The integrated intensity observed by SANS is about 10 times weaker than seen by TAS which relates well to the different divergence. 

Upon approaching the transition at $T_N \approx 3.05$\,K from below, the intensity of the magnetic Bragg peaks decreases by a factor of 4 with respect to coldest temperatures. Moreover, at the critical temperature $T_N$ a sharp drop of the intensity is observed consistent with the notion of a first-order transition. For the ring of intensity, a sharp increase of intensity for temperatures below the transition temperature and an exponential decay for temperatures above the transition is observed as a function of increasing $T$ so that the ring reaches its maximum intensity at $T_N$. The sum of TAS intensity associated with Bragg peaks and the ring (grey points) only shows a small, upward kink at $T_N$, most likely caused by resolution effects. The intensity of the incommensurate AFM ring obtained from TAS is again depicted in panel (B) for three different combinations of the parameters $(K_z,\lambda)$ yielding basically equivalent results. The points for different $(K_z,\lambda)$ are essentially overlapping. 

The temperature dependence of the inverse correlation length $\kappa$, see Eq.~\eqref{kappa}, is displayed in panel (C). Similar to the intensity, the result for $\kappa$ is shown for three different combinations of the parameters $(K_z,\lambda)$, marked by crosses in the FOM (Fig.~\ref{Fig_2}). Again, the points for different $(K_z,\lambda)$ are essentially overlapping. As a function of increasing temperature, the inverse correlation length $\kappa$ decreases rapidly from $0.016/$\AA\, at 2.2\,K to $0.0025/$\AA\, at $T_N$ where an upward kink is observed. Above $T_N$, an exponential growth of $\kappa$ is observed as a function of increasing $T$. 
At the transition, we find the value $\kappa=0.0027/$\AA, that is still well above the instrumental TAS resolution of $0.0015/$\AA. The value $0.0027\pm0.0015/$\AA\, obtained with SANS at $T_N$ is in good agreement with the TAS data.

\section{Summary and Discussion}
\label{sec:summary}
In the following, we summarize and discuss the main results of our neutron diffraction experiments on Ba$_2$CuGe$_2$O$_7$.

\subsection{Weak incommensurate FM order}
The AFM cycloid that forms below the transition temperature $T_N \approx 3.05$\,K was theoretically predicted \cite{Bogdanov:02,Chovan:02} to induce according to Eq.~\eqref{inducedFM} an incommensurate FM density-wave. Our combined TAS and SANS experiments have established that incommensurate Bragg peaks are indeed present close to zero wavevector in addition to the ones close to the N\'{e}el point. In both experiments, the Bragg peaks are resolution limited indicative of long-range magnetic order. As a result, rocking scans are able to capture the complete integrated intensity of the diffraction peaks. Comparing the intensity of the incommensurate AFM and the FM Bragg peaks we found that they are characterized by the same temperature dependence, see Fig.~\ref{Fig_4}(A). Moreover, measurements at different neutron wavelengths show that the Bragg peaks in the SANS data are not caused by double scattering. All in all, this provides strong evidence that the AFM cycloid also induces incommensurate FM long-range order. Above the critical temperature $T_N$, incommensurate short-range FM order is also observed with the same correlation length as in the AFM sector, see Fig.~\ref{Fig_4}(C).

\subsection{Quantitative description of critical AFM and FM fluctuations above $T_N$} 
Above the critical temperature $T_N$, our neutron diffraction data is consistent with magnetic intensity that accumulates on an incommensurate ring in momentum space both around the N\'{e}el point and around the $\Gamma$ point. The ring is located within the plane perpendicular to the tetragonal $c$ axis. The intensity distribution is quantitatively described by the theory of Refs.~\cite{Zheludev:99,Bogdanov:02,Chovan:02} complemented by a term representing magnetocrystalline anisotropies of strength $\lambda$. The data was fitted with two parameters, $\lambda$ and the AFM anisotropy $K_z$. We found a range of value for $(K_z, \lambda)$ that describes equally well our data, see Fig.~\ref{Fig_2}, including the specific value for $K_z$ that corresponds to the KSEA limit \cite{Kaplan1983,Shekhtman1993}. Our neutron diffraction data is thus consistent with the KSEA scenario. The observed ring of intensity implies that the magnetocrystalline anisotropy $\lambda$ is effectively weak.

Our scattering geometry was limited to the crystallographic $(a,b)$ plane. So far, it is therefore not known whether and how the ring of magnetic intensity extends along the $c$ direction in momentum space. As the inter-plane exchange coupling in Ba$_2$CuGe$_2$O$_7$ ($J_{\perp}\approx -0.026\,\rm {meV}$) 
is one order of magnetitude smaller than the intra-plane coupling ($J_{\|}\approx 0.96\,\rm {meV}$ per bond), one expects that the fluctuations above $T_N$ are quasi two-dimensional for an extended range of momenta. This suggests that the ring of intensity extends as a cylinder along the tetragonal $c$ axis. However, due to the small incommesurability and the wide vertical divergence of the TAS experiment of $4-5^{\circ}$ FWHM, an integration regime of almost $\approx 0.07\AA^{-1}$ FWHM essenially covers also fluctuations on a cylinder. The exact extension of fluctuations along the tetragonal axis is has to be confirmed experimentally in a future study in different scattering geometry. This argument is in-line with the weak intensity observed by SANS.

Moreover, also the energy range integrated by a TAS and SANS are different, leading to stronger intensity observed by TAS. Whereas a cold TAS typically integrates over an energy region of $\approx 0.1$\,meV with energy analysis, only the scattering triangle has to be closed for a TAS in two-axis mode and for SANS. However, with a few restrictions on reasonable momentum transfers, an energy window of $\approx \pm0.5$\,meV for TAS is expected. For for diffraction in SANS geometry the enery window is reduced by a factor of 2 due to the better momentum resolution.

\subsection{Weak first-order transition and $T$-dependence of the correlation length}

Our detailed quantitative fit to the experimental diffraction data allows us to determine the temperature dependence of the inverse correlation length $\kappa$ of the short-range incommensurate magnetic order above the critical temperature $T_N$, see Fig.~\ref{Fig_4}(C). As $T_N$ is approached from above, the correlation length $1/\kappa$ increases and assumes a finite value $1/\kappa_{\rm cr} = (1/0.0027)$\AA\, at the transition corresponding to $1/(\kappa_{\rm cr} a) \approx 44$ lattice units or $k_{\rm cycl.}/(2\pi \kappa_{\rm cr}) \approx 1.6$ wavelengths of the incommensurate magnetic order. The finite value of $\kappa_{\rm cr}$ together with the sudden rise of intensity associated with the magnetic Bragg peak below $T_N$, see Fig.~\ref{Fig_4}(A), confirms the weak first-order character of the magnetic phase transition. 

It is suggestive that this weak first-order transition in Ba$_2$CuGe$_2$O$_7$ is explained by the Brazovskii scenario, i.e., that it is induced by the critical fluctuations becoming soft on a ring in momentum space. The question then arises whether Brazovskii theory \cite{Brazovskii:1975wr} can also account for the temperature dependence of $\kappa$. In the cubic helimagnet MnSi, the inverse correlation length was found to exhibit a crossover from a concave, mean-field like behavior, $\partial_T^2 \kappa < 0$, at higher $T$ to a convex behavior, $\partial_T^2 \kappa > 0$, close to the critical temperature \cite{Janosch:13}. This crossover in MnSi is well described by Brazovskii theory.

In Ba$_2$CuGe$_2$O$_7$, we find a $T$-dependence of $\kappa$ that is convex, $\partial_T^2 \kappa > 0$, within the investigated temperature range above $T_N$, see Fig.~\ref{Fig_4}(C). This excludes, in particular, the mean-field behavior $\kappa \sim \sqrt{T-T_N}$. It is well possible that the crossover temperature $T^*$, beyond which mean-field behavior is recovered, is larger than the accessible temperatures of Fig.~\ref{Fig_4}(C). We tried to fit the $T$-dependence of $\kappa$ with Brazovskii theory for $T \ll T^*$ but we found that it does not provide a satisfactory description. Instead, we find that it is better described by an exponential dependence 
\begin{align} \label{FitKappa}
\kappa = \kappa_0\, e^{-\Delta/{k_B T}}
\end{align}
with the fitting parameters $\kappa_0 = 29.1/$\AA\,  and $\Delta/k_B = 28.6$ K corresponding to $\Delta/J_\parallel \approx 2.6$. This fit is shown as a solid green line in Fig.~\ref{Fig_4}(C).

While the Brazovskii scenario might account for the first-order character of the transition in Ba$_2$CuGe$_2$O$_7$, the approximation involved in the Brazovskii theory might just not be applicable. Brazovskii theory is a non-perturbative description of the fluctuation-induced first-order transition that is controlled in the limit of a small amplitude for the self-interaction of the fluctuations. It accounts for the non-perturbative renormalization from interacting critical fluctuations with a wavelength {\it larger} than the wavelength of the cycloid, $2\pi/k_{\rm cycl.}$~provided that the impact from fluctuations with smaller wavelength is perturbative. 
However, if the magnetism possesses a two-dimensional character like in tetragonal Ba$_2$CuGe$_2$O$_7$ the latter assumption could break down; the interaction amplitude gets already enhanced by exactly those fluctuations with a wavelength {\it smaller} than the wavelength of the cycloid, $2\pi/k_{\rm cycl.}$. This might invalidate the applicability of the Brazovskii approximation. In fact, for such small wavelengths the theory of Eq.~\eqref{theory} effectively reduces to the low-energy description of the two-dimensional AFM Heisenberg model, whose correlation length possesses a $T$-dependence of the form given in Eq.~\eqref{FitKappa} \cite{HASENFRATZ1991}. 
Before long-range order develops, the Dzyaloshinskii-Moriya interaction drives a crossover from a regime at higher $T$ to a strongly correlated chiral paramagnet. Whereas in MnSi the regime at higher $T$ is captured by mean-field behavior, our findings suggest that in Ba$_2$CuGe$_2$O$_7$ the regime at higher $T$ is instead governed by the non-perturbative physics of the two-dimensional AFM Heisenberg model. Whereas the physics above the critical temperature is dominated by two-dimensional fluctuations, the magnetic order that eventually develops below $T_N$ is  three-dimensional due to the weak interlayer coupling $J_\perp$.

\section{Acknowledgments}

We wish to thank E. Pomjakushina, K. Conder, P. B\"oni, C. Pfleiderer and A. Zheludev for support and stimulating discussions. Technical support and help from M. Bartkowiak, M. Zolliker, A. Wilhelm, S. Semecky and H. Kolb is gratefully acknowledged. M.G. acknowledges support from DFG SFB1143 (Correlated Magnetism: From Frustration To Topology) and DFG Grant GA 1072/5-1. M.M. acknowledges support from the Swedish Research Council (VR) through a project research grant in neutron scattering (Dnr 2016-06955).


\begin{thebibliography}{34}
\expandafter\ifx\csname natexlab\endcsname\relax\def\natexlab#1{#1}\fi
\expandafter\ifx\csname bibnamefont\endcsname\relax
  \def\bibnamefont#1{#1}\fi
\expandafter\ifx\csname bibfnamefont\endcsname\relax
  \def\bibfnamefont#1{#1}\fi
\expandafter\ifx\csname citenamefont\endcsname\relax
  \def\citenamefont#1{#1}\fi
\expandafter\ifx\csname url\endcsname\relax
  \def\url#1{\texttt{#1}}\fi
\expandafter\ifx\csname urlprefix\endcsname\relax\def\urlprefix{URL }\fi
\providecommand{\bibinfo}[2]{#2}
\providecommand{\eprint}[2][]{\url{#2}}

\bibitem[{\citenamefont{Dzyaloshinskii}(1958)}]{Dzyaloshinskii:58}
\bibinfo{author}{\bibfnamefont{I.~E.} \bibnamefont{Dzyaloshinskii}},
  \bibinfo{journal}{J. Phys. Chem Solids} \textbf{\bibinfo{volume}{4}},
  \bibinfo{pages}{241} (\bibinfo{year}{1958}).

\bibitem[{\citenamefont{Moriya}(1960)}]{Moriya:60}
\bibinfo{author}{\bibfnamefont{T.}~\bibnamefont{Moriya}},
  \bibinfo{journal}{Phys. Rev.} \textbf{\bibinfo{volume}{120}},
  \bibinfo{pages}{91} (\bibinfo{year}{1960}).

\bibitem[{\citenamefont{Brazovskii}(1975)}]{Brazovskii:1975wr}
\bibinfo{author}{\bibfnamefont{S.~A.} \bibnamefont{Brazovskii}},
  \bibinfo{journal}{Sov. Phys. JETP} \textbf{\bibinfo{volume}{41}},
  \bibinfo{pages}{85} (\bibinfo{year}{1975}).

\bibitem[{\citenamefont{Brazovskii et~al.}(1987)\citenamefont{Brazovskii,
  Dzyaloshinskii, and Muratov}}]{1987ZhETF..93.1110B}
\bibinfo{author}{\bibfnamefont{S.~A.} \bibnamefont{Brazovskii}},
  \bibinfo{author}{\bibfnamefont{I.~E.} \bibnamefont{Dzyaloshinskii}},
  \bibnamefont{and} \bibinfo{author}{\bibfnamefont{A.~R.}
  \bibnamefont{Muratov}}, \bibinfo{journal}{Sov. Phys. JETP}
  \textbf{\bibinfo{volume}{66}}, \bibinfo{pages}{625} (\bibinfo{year}{1987}).

\bibitem[{\citenamefont{Brazovskii and Dmitriev}(1975)}]{1975JETP...42..497B}
\bibinfo{author}{\bibfnamefont{S.~A.} \bibnamefont{Brazovskii}}
  \bibnamefont{and} \bibinfo{author}{\bibfnamefont{S.~G.}
  \bibnamefont{Dmitriev}}, \bibinfo{journal}{Sov. Phys. JETP}
  \textbf{\bibinfo{volume}{42}}, \bibinfo{pages}{497} (\bibinfo{year}{1975}).

\bibitem[{\citenamefont{Swift}(1976)}]{Swift:1976vp}
\bibinfo{author}{\bibfnamefont{J.}~\bibnamefont{Swift}},
  \bibinfo{journal}{Phys. Rev. A} \textbf{\bibinfo{volume}{14}},
  \bibinfo{pages}{2274} (\bibinfo{year}{1976}).

\bibitem[{\citenamefont{Leibler}(1980)}]{Leibler:1980ti}
\bibinfo{author}{\bibfnamefont{L.}~\bibnamefont{Leibler}},
  \bibinfo{journal}{Macromolecules} \textbf{\bibinfo{volume}{13}},
  \bibinfo{pages}{1602} (\bibinfo{year}{1980}).

\bibitem[{\citenamefont{Bates et~al.}(1990)\citenamefont{Bates, Rosedale, and
  Fredrickson}}]{Bates:1990hi}
\bibinfo{author}{\bibfnamefont{F.~S.} \bibnamefont{Bates}},
  \bibinfo{author}{\bibfnamefont{J.~H.} \bibnamefont{Rosedale}},
  \bibnamefont{and} \bibinfo{author}{\bibfnamefont{G.~H.}
  \bibnamefont{Fredrickson}}, \bibinfo{journal}{J. Chem. Phys.}
  \textbf{\bibinfo{volume}{92}}, \bibinfo{pages}{6255} (\bibinfo{year}{1990}).

\bibitem[{\citenamefont{Migdal et~al.}(1990)\citenamefont{Migdal, Saperstein,
  Troitsky, and Voskresensky}}]{1990PhR...192..179M}
\bibinfo{author}{\bibfnamefont{A.~B.} \bibnamefont{Migdal}},
  \bibinfo{author}{\bibfnamefont{E.~E.} \bibnamefont{Saperstein}},
  \bibinfo{author}{\bibfnamefont{M.~A.} \bibnamefont{Troitsky}},
  \bibnamefont{and} \bibinfo{author}{\bibfnamefont{D.~N.}
  \bibnamefont{Voskresensky}}, \bibinfo{journal}{Phys. Rep.}
  \textbf{\bibinfo{volume}{192}}, \bibinfo{pages}{179} (\bibinfo{year}{1990}).

\bibitem[{\citenamefont{Gopalakrishnan
  et~al.}(2009)\citenamefont{Gopalakrishnan, Lev, and
  Goldbart}}]{Gopalakrishnan:2009gv}
\bibinfo{author}{\bibfnamefont{S.}~\bibnamefont{Gopalakrishnan}},
  \bibinfo{author}{\bibfnamefont{B.~L.} \bibnamefont{Lev}}, \bibnamefont{and}
  \bibinfo{author}{\bibfnamefont{P.~M.} \bibnamefont{Goldbart}},
  \bibinfo{journal}{Nat. Phys.} \textbf{\bibinfo{volume}{5}},
  \bibinfo{pages}{845} (\bibinfo{year}{2009}).

\bibitem[{\citenamefont{Janoschek et~al.}(2013)\citenamefont{Janoschek, Garst,
  Bauer, Krautscheid, Georgii, B\"oni, and Pfleiderer}}]{Janosch:13}
\bibinfo{author}{\bibfnamefont{M.}~\bibnamefont{Janoschek}},
  \bibinfo{author}{\bibfnamefont{M.}~\bibnamefont{Garst}},
  \bibinfo{author}{\bibfnamefont{A.}~\bibnamefont{Bauer}},
  \bibinfo{author}{\bibfnamefont{P.}~\bibnamefont{Krautscheid}},
  \bibinfo{author}{\bibfnamefont{R.}~\bibnamefont{Georgii}},
  \bibinfo{author}{\bibfnamefont{P.}~\bibnamefont{B\"oni}}, \bibnamefont{and}
  \bibinfo{author}{\bibfnamefont{C.}~\bibnamefont{Pfleiderer}},
  \bibinfo{journal}{Phys. Rev. B} \textbf{\bibinfo{volume}{87}},
  \bibinfo{pages}{134407} (\bibinfo{year}{2013}).

\bibitem[{\citenamefont{Stishov et~al.}(2007)\citenamefont{Stishov, Petrova,
  Khasanov, Panova, Shikov, Lashley, Wu, and Lograsso}}]{Stishov:2007jf}
\bibinfo{author}{\bibfnamefont{S.}~\bibnamefont{Stishov}},
  \bibinfo{author}{\bibfnamefont{A.}~\bibnamefont{Petrova}},
  \bibinfo{author}{\bibfnamefont{S.}~\bibnamefont{Khasanov}},
  \bibinfo{author}{\bibfnamefont{G.}~\bibnamefont{Panova}},
  \bibinfo{author}{\bibfnamefont{A.}~\bibnamefont{Shikov}},
  \bibinfo{author}{\bibfnamefont{J.}~\bibnamefont{Lashley}},
  \bibinfo{author}{\bibfnamefont{D.}~\bibnamefont{Wu}}, \bibnamefont{and}
  \bibinfo{author}{\bibfnamefont{T.}~\bibnamefont{Lograsso}},
  \bibinfo{journal}{Phys. Rev. B} \textbf{\bibinfo{volume}{76}}
  (\bibinfo{year}{2007}).

\bibitem[{\citenamefont{Bauer et~al.}(2013)\citenamefont{Bauer, Garst, and
  Pfleiderer}}]{2013PhRvL.110q7207B}
\bibinfo{author}{\bibfnamefont{A.}~\bibnamefont{Bauer}},
  \bibinfo{author}{\bibfnamefont{M.}~\bibnamefont{Garst}}, \bibnamefont{and}
  \bibinfo{author}{\bibfnamefont{C.}~\bibnamefont{Pfleiderer}},
  \bibinfo{journal}{Phys. Rev. Lett.} \textbf{\bibinfo{volume}{110}},
  \bibinfo{pages}{177207} (\bibinfo{year}{2013}).

\bibitem[{\citenamefont{\ifmmode \check{Z}\else
  \v{Z}\fi{}ivkovi\ifmmode~\acute{c}\else \'{c}\fi{}
  et~al.}(2014)\citenamefont{\ifmmode \check{Z}\else
  \v{Z}\fi{}ivkovi\ifmmode~\acute{c}\else \'{c}\fi{}, White, R\o{}nnow,
  Pr\ifmmode~\check{s}\else \v{s}\fi{}a, and Berger}}]{PhysRevB.89.060401}
\bibinfo{author}{\bibfnamefont{I.}~\bibnamefont{\ifmmode \check{Z}\else
  \v{Z}\fi{}ivkovi\ifmmode~\acute{c}\else \'{c}\fi{}}},
  \bibinfo{author}{\bibfnamefont{J.~S.} \bibnamefont{White}},
  \bibinfo{author}{\bibfnamefont{H.~M.} \bibnamefont{R\o{}nnow}},
  \bibinfo{author}{\bibfnamefont{K.}~\bibnamefont{Pr\ifmmode~\check{s}\else
  \v{s}\fi{}a}}, \bibnamefont{and}
  \bibinfo{author}{\bibfnamefont{H.}~\bibnamefont{Berger}},
  \bibinfo{journal}{Phys. Rev. B} \textbf{\bibinfo{volume}{89}},
  \bibinfo{pages}{060401} (\bibinfo{year}{2014}).

\bibitem[{\citenamefont{{Bannenberg} et~al.}(2017)\citenamefont{{Bannenberg},
  {Dalgliesh}, {Falus}, {Leli{\`e}vre-Berna}, {Dewhurst}, {Qian}, {Onose},
  {Endoh}, {Tokura}, {Kakurai} et~al.}}]{2017arXiv170105448B}
\bibinfo{author}{\bibfnamefont{L.~J.} \bibnamefont{{Bannenberg}}},
  \bibinfo{author}{\bibfnamefont{R.}~\bibnamefont{{Dalgliesh}}},
  \bibinfo{author}{\bibfnamefont{P.}~\bibnamefont{{Falus}}},
  \bibinfo{author}{\bibfnamefont{E.}~\bibnamefont{{Leli{\`e}vre-Berna}}},
  \bibinfo{author}{\bibfnamefont{C.~D.} \bibnamefont{{Dewhurst}}},
  \bibinfo{author}{\bibfnamefont{F.}~\bibnamefont{{Qian}}},
  \bibinfo{author}{\bibfnamefont{Y.}~\bibnamefont{{Onose}}},
  \bibinfo{author}{\bibfnamefont{Y.}~\bibnamefont{{Endoh}}},
  \bibinfo{author}{\bibfnamefont{Y.}~\bibnamefont{{Tokura}}},
  \bibinfo{author}{\bibfnamefont{K.}~\bibnamefont{{Kakurai}}},
  \bibnamefont{et~al.}, \bibinfo{journal}{ArXiv e-prints}
  (\bibinfo{year}{2017}), \eprint{1701.05448}.

\bibitem[{\citenamefont{Zheludev et~al.}(1999)\citenamefont{Zheludev, Maslov,
  Shirane, Tsukada, Masuda, Uchinokura, Zaliznyak, Erwin, and
  Regnault}}]{Zheludev:99}
\bibinfo{author}{\bibfnamefont{A.}~\bibnamefont{Zheludev}},
  \bibinfo{author}{\bibfnamefont{S.}~\bibnamefont{Maslov}},
  \bibinfo{author}{\bibfnamefont{G.}~\bibnamefont{Shirane}},
  \bibinfo{author}{\bibfnamefont{I.}~\bibnamefont{Tsukada}},
  \bibinfo{author}{\bibfnamefont{T.}~\bibnamefont{Masuda}},
  \bibinfo{author}{\bibfnamefont{K.}~\bibnamefont{Uchinokura}},
  \bibinfo{author}{\bibfnamefont{I.}~\bibnamefont{Zaliznyak}},
  \bibinfo{author}{\bibfnamefont{R.}~\bibnamefont{Erwin}}, \bibnamefont{and}
  \bibinfo{author}{\bibfnamefont{L.~P.} \bibnamefont{Regnault}},
  \bibinfo{journal}{Phys. Rev. B} \textbf{\bibinfo{volume}{59}},
  \bibinfo{pages}{11432} (\bibinfo{year}{1999}).

\bibitem[{\citenamefont{M\"uhlbauer et~al.}(2012)\citenamefont{M\"uhlbauer,
  Gvasaliya, Ressouche, Pomjakushina, and Zheludev}}]{muehlbauer:12}
\bibinfo{author}{\bibfnamefont{S.}~\bibnamefont{M\"uhlbauer}},
  \bibinfo{author}{\bibfnamefont{S.}~\bibnamefont{Gvasaliya}},
  \bibinfo{author}{\bibfnamefont{E.}~\bibnamefont{Ressouche}},
  \bibinfo{author}{\bibfnamefont{E.}~\bibnamefont{Pomjakushina}},
  \bibnamefont{and} \bibinfo{author}{\bibfnamefont{A.}~\bibnamefont{Zheludev}},
  \bibinfo{journal}{Phys. Rev. B} \textbf{\bibinfo{volume}{86}},
  \bibinfo{pages}{024417} (\bibinfo{year}{2012}).

\bibitem[{\citenamefont{Zheludev et~al.}(1996)\citenamefont{Zheludev, Shirane,
  Sasago, Kiode, and Uchinokura}}]{Zheludev:96}
\bibinfo{author}{\bibfnamefont{A.}~\bibnamefont{Zheludev}},
  \bibinfo{author}{\bibfnamefont{G.}~\bibnamefont{Shirane}},
  \bibinfo{author}{\bibfnamefont{Y.}~\bibnamefont{Sasago}},
  \bibinfo{author}{\bibfnamefont{N.}~\bibnamefont{Kiode}}, \bibnamefont{and}
  \bibinfo{author}{\bibfnamefont{K.}~\bibnamefont{Uchinokura}},
  \bibinfo{journal}{Phys. Rev. B} \textbf{\bibinfo{volume}{54}},
  \bibinfo{pages}{15163} (\bibinfo{year}{1996}).

\bibitem[{\citenamefont{Zheludev
  et~al.}(1997{\natexlab{a}})\citenamefont{Zheludev, Maslov, Shirane, Sasago,
  Koide, and Uchinokura}}]{Zheludev:97}
\bibinfo{author}{\bibfnamefont{A.}~\bibnamefont{Zheludev}},
  \bibinfo{author}{\bibfnamefont{S.}~\bibnamefont{Maslov}},
  \bibinfo{author}{\bibfnamefont{G.}~\bibnamefont{Shirane}},
  \bibinfo{author}{\bibfnamefont{Y.}~\bibnamefont{Sasago}},
  \bibinfo{author}{\bibfnamefont{N.}~\bibnamefont{Koide}}, \bibnamefont{and}
  \bibinfo{author}{\bibfnamefont{K.}~\bibnamefont{Uchinokura}},
  \bibinfo{journal}{Phys. Rev. Lett.} \textbf{\bibinfo{volume}{78}},
  \bibinfo{pages}{4857} (\bibinfo{year}{1997}{\natexlab{a}}).

\bibitem[{\citenamefont{Zheludev
  et~al.}(1997{\natexlab{b}})\citenamefont{Zheludev, Maslov, Shirane, Sasago,
  Koide, Uchinokura, Tennant, and Nagler}}]{Zheludev:97b}
\bibinfo{author}{\bibfnamefont{A.}~\bibnamefont{Zheludev}},
  \bibinfo{author}{\bibfnamefont{S.}~\bibnamefont{Maslov}},
  \bibinfo{author}{\bibfnamefont{G.}~\bibnamefont{Shirane}},
  \bibinfo{author}{\bibfnamefont{Y.}~\bibnamefont{Sasago}},
  \bibinfo{author}{\bibfnamefont{N.}~\bibnamefont{Koide}},
  \bibinfo{author}{\bibfnamefont{K.}~\bibnamefont{Uchinokura}},
  \bibinfo{author}{\bibfnamefont{D.~A.} \bibnamefont{Tennant}},
  \bibnamefont{and} \bibinfo{author}{\bibfnamefont{S.~E.}
  \bibnamefont{Nagler}}, \bibinfo{journal}{Phys. Rev. B}
  \textbf{\bibinfo{volume}{56}}, \bibinfo{pages}{14006}
  (\bibinfo{year}{1997}{\natexlab{b}}).

\bibitem[{\citenamefont{Zheludev
  et~al.}(1997{\natexlab{c}})\citenamefont{Zheludev, Shirane, Sasago, Koide,
  and Uchinokura}}]{Zheludev:97PB}
\bibinfo{author}{\bibfnamefont{A.}~\bibnamefont{Zheludev}},
  \bibinfo{author}{\bibfnamefont{G.}~\bibnamefont{Shirane}},
  \bibinfo{author}{\bibfnamefont{Y.}~\bibnamefont{Sasago}},
  \bibinfo{author}{\bibfnamefont{N.}~\bibnamefont{Koide}}, \bibnamefont{and}
  \bibinfo{author}{\bibfnamefont{K.}~\bibnamefont{Uchinokura}},
  \bibinfo{journal}{Physica B: Condensed Matter}
  \textbf{\bibinfo{volume}{234–}}, \bibinfo{pages}{546 }
  (\bibinfo{year}{1997}{\natexlab{c}}).

\bibitem[{\citenamefont{M\"uhlbauer et~al.}(2011)\citenamefont{M\"uhlbauer,
  Gvasaliya, Pomjakushina, and Zheludev}}]{Muehlbauer:11}
\bibinfo{author}{\bibfnamefont{S.}~\bibnamefont{M\"uhlbauer}},
  \bibinfo{author}{\bibfnamefont{S.~N.} \bibnamefont{Gvasaliya}},
  \bibinfo{author}{\bibfnamefont{E.}~\bibnamefont{Pomjakushina}},
  \bibnamefont{and} \bibinfo{author}{\bibfnamefont{A.}~\bibnamefont{Zheludev}},
  \bibinfo{journal}{Phys. Rev. B} \textbf{\bibinfo{volume}{84}},
  \bibinfo{pages}{180406} (\bibinfo{year}{2011}).

\bibitem[{\citenamefont{Bogdanov et~al.}(2002)\citenamefont{Bogdanov,
  R\"o\ss{}ler, Wolf, and M\"uller}}]{Bogdanov:02}
\bibinfo{author}{\bibfnamefont{A.~N.} \bibnamefont{Bogdanov}},
  \bibinfo{author}{\bibfnamefont{U.~K.} \bibnamefont{R\"o\ss{}ler}},
  \bibinfo{author}{\bibfnamefont{M.}~\bibnamefont{Wolf}}, \bibnamefont{and}
  \bibinfo{author}{\bibfnamefont{K.-H.} \bibnamefont{M\"uller}},
  \bibinfo{journal}{Phys. Rev. B} \textbf{\bibinfo{volume}{66}},
  \bibinfo{pages}{214410} (\bibinfo{year}{2002}).

\bibitem[{\citenamefont{Chovan et~al.}(2002)\citenamefont{Chovan, Papanicolaou,
  and Komineas}}]{Chovan:02}
\bibinfo{author}{\bibfnamefont{J.}~\bibnamefont{Chovan}},
  \bibinfo{author}{\bibfnamefont{N.}~\bibnamefont{Papanicolaou}},
  \bibnamefont{and} \bibinfo{author}{\bibfnamefont{S.}~\bibnamefont{Komineas}},
  \bibinfo{journal}{Phys. Rev. B} \textbf{\bibinfo{volume}{65}},
  \bibinfo{pages}{064433} (\bibinfo{year}{2002}).

\bibitem[{\citenamefont{Chovan et~al.}(2013)\citenamefont{Chovan, Marder, and
  Papanicolaou}}]{Chovan:13}
\bibinfo{author}{\bibfnamefont{J.}~\bibnamefont{Chovan}},
  \bibinfo{author}{\bibfnamefont{M.}~\bibnamefont{Marder}}, \bibnamefont{and}
  \bibinfo{author}{\bibfnamefont{N.}~\bibnamefont{Papanicolaou}},
  \bibinfo{journal}{Phys. Rev. B} \textbf{\bibinfo{volume}{88}},
  \bibinfo{pages}{064421} (\bibinfo{year}{2013}).

\bibitem[{\citenamefont{Bauer et~al.}(2010)\citenamefont{Bauer, Neubauer,
  Franz, M\"unzer, Garst, and Pfleiderer}}]{Bauer:10}
\bibinfo{author}{\bibfnamefont{A.}~\bibnamefont{Bauer}},
  \bibinfo{author}{\bibfnamefont{A.}~\bibnamefont{Neubauer}},
  \bibinfo{author}{\bibfnamefont{C.}~\bibnamefont{Franz}},
  \bibinfo{author}{\bibfnamefont{W.}~\bibnamefont{M\"unzer}},
  \bibinfo{author}{\bibfnamefont{M.}~\bibnamefont{Garst}}, \bibnamefont{and}
  \bibinfo{author}{\bibfnamefont{C.}~\bibnamefont{Pfleiderer}},
  \bibinfo{journal}{Phys. Rev. B} \textbf{\bibinfo{volume}{82}},
  \bibinfo{pages}{064404} (\bibinfo{year}{2010}).

\bibitem[{\citenamefont{Kaplan}(1983)}]{Kaplan1983}
\bibinfo{author}{\bibfnamefont{T.~A.} \bibnamefont{Kaplan}},
  \bibinfo{journal}{Z. Phys. B: Condens. Matter} \textbf{\bibinfo{volume}{49}},
  \bibinfo{pages}{313} (\bibinfo{year}{1983}).

\bibitem[{\citenamefont{Shekhtman et~al.}(1993)\citenamefont{Shekhtman,
  Aharony, and Entin-Wohlman}}]{Shekhtman1993}
\bibinfo{author}{\bibfnamefont{L.}~\bibnamefont{Shekhtman}},
  \bibinfo{author}{\bibfnamefont{A.}~\bibnamefont{Aharony}}, \bibnamefont{and}
  \bibinfo{author}{\bibfnamefont{O.}~\bibnamefont{Entin-Wohlman}},
  \bibinfo{journal}{Phys. Rev. B} \textbf{\bibinfo{volume}{47}},
  \bibinfo{pages}{174} (\bibinfo{year}{1993}).

\bibitem[{\citenamefont{M\"uhlbauer et~al.}(2015)\citenamefont{M\"uhlbauer,
  Kindervater, and H\"au\ss{}ler}}]{Muehlbauer:15}
\bibinfo{author}{\bibfnamefont{S.}~\bibnamefont{M\"uhlbauer}},
  \bibinfo{author}{\bibfnamefont{J.}~\bibnamefont{Kindervater}},
  \bibnamefont{and}
  \bibinfo{author}{\bibfnamefont{W.}~\bibnamefont{H\"au\ss{}ler}},
  \bibinfo{journal}{Phys. Rev. B} \textbf{\bibinfo{volume}{92}},
  \bibinfo{pages}{224406} (\bibinfo{year}{2015}).

\bibitem[{\citenamefont{Semadeni et~al.}(2001)\citenamefont{Semadeni, Roessli,
  and B{\"o}ni}}]{semadeni:01}
\bibinfo{author}{\bibfnamefont{F.}~\bibnamefont{Semadeni}},
  \bibinfo{author}{\bibfnamefont{B.}~\bibnamefont{Roessli}}, \bibnamefont{and}
  \bibinfo{author}{\bibfnamefont{P.}~\bibnamefont{B{\"o}ni}},
  \bibinfo{journal}{Physica B} \textbf{\bibinfo{volume}{297}},
  \bibinfo{pages}{152} (\bibinfo{year}{2001}).

\bibitem[{\citenamefont{Janoschek et~al.}(2007)\citenamefont{Janoschek, Klimko,
  Roessli, G{\"a}hler, and B{\"o}ni}}]{janoschek:07}
\bibinfo{author}{\bibfnamefont{M.}~\bibnamefont{Janoschek}},
  \bibinfo{author}{\bibfnamefont{S.}~\bibnamefont{Klimko}},
  \bibinfo{author}{\bibfnamefont{B.}~\bibnamefont{Roessli}},
  \bibinfo{author}{\bibfnamefont{R.}~\bibnamefont{G{\"a}hler}},
  \bibnamefont{and} \bibinfo{author}{\bibfnamefont{P.}~\bibnamefont{B{\"o}ni}},
  \bibinfo{journal}{Physica B} \textbf{\bibinfo{volume}{397}},
  \bibinfo{pages}{125} (\bibinfo{year}{2007}).

\bibitem[{\citenamefont{{Heinz Maier Leibnitz Zentrum et al.}}(2015)}]{SANS-1}
\bibinfo{author}{\bibnamefont{{Heinz Maier Leibnitz Zentrum et al.}}},
  \bibinfo{journal}{Journal of large-scale research facilities}
  \textbf{\bibinfo{volume}{1}} (\bibinfo{year}{2015}).

\bibitem[{\citenamefont{M{\"u}hlbauer et~al.}(2016)\citenamefont{M{\"u}hlbauer,
  Heinemann, Wilhelm, Karge, Ostermann, Defendi, Schreyer, Petry, and
  Gilles}}]{Muehlbauer:16}
\bibinfo{author}{\bibfnamefont{S.}~\bibnamefont{M{\"u}hlbauer}},
  \bibinfo{author}{\bibfnamefont{A.}~\bibnamefont{Heinemann}},
  \bibinfo{author}{\bibfnamefont{A.}~\bibnamefont{Wilhelm}},
  \bibinfo{author}{\bibfnamefont{L.}~\bibnamefont{Karge}},
  \bibinfo{author}{\bibfnamefont{A.}~\bibnamefont{Ostermann}},
  \bibinfo{author}{\bibfnamefont{I.}~\bibnamefont{Defendi}},
  \bibinfo{author}{\bibfnamefont{A.}~\bibnamefont{Schreyer}},
  \bibinfo{author}{\bibfnamefont{W.}~\bibnamefont{Petry}}, \bibnamefont{and}
  \bibinfo{author}{\bibfnamefont{R.}~\bibnamefont{Gilles}},
  \bibinfo{journal}{Nuclear Instruments and Methods in Physics Research Section
  A: Accelerators, Spectrometers, Detectors and Associated Equipment}
  \textbf{\bibinfo{volume}{832}}, \bibinfo{pages}{297} (\bibinfo{year}{2016}).

\bibitem[{\citenamefont{Hasenfratz and Niedermayer}(1991)}]{HASENFRATZ1991}
\bibinfo{author}{\bibfnamefont{P.}~\bibnamefont{Hasenfratz}} \bibnamefont{and}
  \bibinfo{author}{\bibfnamefont{F.}~\bibnamefont{Niedermayer}},
  \bibinfo{journal}{Physics Letters B} \textbf{\bibinfo{volume}{268}},
  \bibinfo{pages}{231} (\bibinfo{year}{1991}).

\end{thebibliography}
\end{document}